\documentclass[pre,twocolumn,showpacs,amsmath,amssymb]{revtex4}
\usepackage[latin1]{inputenc}
\usepackage{amssymb,amsmath}
\usepackage{graphicx}

\def\LM#1#2{\left|\begin{array}{l}{#1}\\[1ex]{#2}\end{array}\right.}

\begin{document}
\title{Average shape of  fluctuations for subdiffusive walks}
\author{S. B. Yuste}
\email[E-mail: ]{santos@unex.es}
\author{L. Acedo}
\email[E-mail: ]{acedo@unex.es}
\affiliation{%
Departamento de F\'{\i}sica, Universidad  de  Extremadura, E-06071
 Badajoz, Spain
}%

\date{\today}
\begin{abstract}
We study the average shape of fluctuations for subdiffusive
processes, i.e., processes with uncorrelated increments but where
the waiting time distribution has a broad power-law tail. This
shape is obtained analytically by means of a fractional diffusion
approach. We find that, in contrast with processes where the
waiting time between increments has finite variance, the
fluctuation shape is no longer a semicircle: it tends to adopt a
table-like form as the subdiffusive character of the process
increases. The theoretical predictions are compared with numerical
simulation results.

\end{abstract}
 \maketitle

\section{Introduction}

Complex systems are often described by their lack of a
characteristic length or time scale over many orders of magnitude,
which gives rise to events whose distribution in sizes is a power
law with no characteristic size (fractal behaviour). Examples are
everywhere \cite{FractalBooks}: seismic activity, turbulence,
solar flares, Brownian motion, length of rivers and blood
vessels,\ldots In particular, a power-law distribution with no
characteristic \emph{temporal} size events appears in the analysis
of stock price changes\cite{BouchaudCM00,BakPhysicaA97}, river
floods \cite{FederBookFractals} , Barkhausen noise
\cite{KuntzSethnaPRB00,SpasojevicPRE96}, glassy systems
\cite{BouchaudJPI,ArousPRL02,BertinPRE03}, atomic cooling
\cite{SaubameaPRL99}, and fluorescence of quantum dots
\cite{ShimizuPRB01,BrokmannConMat02}. In these cases the resulting
dynamics is strongly intermittent, with bursts of activity
separated by long quiescent intervals.

\begin{figure}[b]
\includegraphics[width=0.77 \columnwidth]{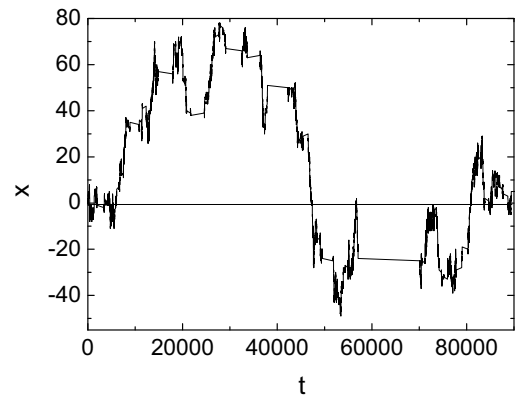}
\caption{Subdiffusive trajectory with $\delta x=\pm 1$ and $\gamma=0.9$.
\label{xtga09.eps}}
\end{figure}

When these temporal intervals are waiting times $\delta t$ between
jumps of size $\delta x$, then the stochastic process $x(t)$ can
be seen as the trajectory of a subdiffusive random walker
(provided that the variance of $\delta x$ is finite). A typical
subdiffusive trajectory is shown in Fig.\ \ref{xtga09.eps}.
Systems that exhibit anomalous subdiffusion characterized by an
anomalous Fick's second law
\begin{equation}\label{FickLaw}
\left< x^2(t)\right> \sim \frac{2K_\gamma}{\Gamma(1+\gamma)}
t^\gamma ,
\end{equation}
where $0<\gamma <1$, are ubiquitous in nature
\cite{MetKlaPhysRep,BouchaudPhysRep90}. [$K_\gamma$ is the
(generalized) diffusion constant and $\gamma$ is the anomalous
diffusion exponent.] Also, models based on subdiffusive random
walkers are useful for understanding complex systems. Two nice
examples are the ``trap model'', proposed to explain aging in
disordered systems \cite{BouchaudJPI,ArousPRL02,BertinPRE03}, and
the comb model, to understand  diffusion phenomena in complex
 structures such as percolation clusters
\cite{HavlinAvrahamAdvPhys87}.

Recently
\cite{SpasojevicPRE96,KuntzSethnaPRB00,BaldassarriPRL03,SethnaNature01}
the study of the average shape of the fluctuations of stochastic
processes $x(t)$  has been considered as a useful tool to gain
insight into the system that generates the fluctuation. Thus, it
has been argued \cite{SethnaNature01} that the average shape of
fluctuations is a better tool for discriminating between theories
than critical exponents.

In Ref.\ \cite{BaldassarriPRL03} Baldassarri \textit{et al.}  consider the
average shape for a stochastic process of the form $x(t+1)=x(t)+\delta x$,
where $\delta x$ is a random variable. Let us denote the average shape of
fluctuations of time span $T$ by $\langle x(t) \rangle_T$.   Baldassarri
\textit{et al.} find that the shape follows the scaling law
\begin{equation}\label{}
\langle x(t) \rangle_T=T^{\alpha/2}f(t/T),
\end{equation}
where $\alpha$ is the diffusion exponent and $f$ is a
\emph{semicircle}  whenever $\delta x$ follows a distribution with
finite variance (Gaussian walks) or a distribution $\lambda(\delta
x)$ with a broad power-law tail, $\lambda(\delta x) \sim (\delta
x)^{-\mu-1}$ with $0<\mu<2$, so that the variance is infinite
(L\'evy flights). This result had already been obtained  by Fisher
for Gaussian walks \cite[Sec.~7.1]{FisherJSP84}. The fact that the
average shape of fluctuations is a semicircle for both Gaussian
walks and L\'evy flights is a nicely surprising result that led us
to wonder to what extent it might hold for other walks with
uncorrelated jumps. In particular, we investigated the average
shape of subdiffusive stochastic processes (subdiffusive walks) of
the form
\begin{equation}\label{xtstosub}
x(t+\delta t)=x(t)+\delta x
\end{equation}
where $\delta x$ are uncorrelated increments that follow a distribution with
finite variance and $\delta t$ is a random variable whose distribution
$\psi(\delta t)$ has a broad power-law tail: $\psi(\delta t) \sim (\delta
t)^{-1-\gamma}$ with $0<\gamma<1$. Figure~\ref{xtga09.eps} shows the position
$x(t)$ for a stochastic process of this kind with $\gamma=0.9$ and with jumps
of  unit length ($\delta=\pm 1$) made with equal probabilities.

\section{Average form of a subdiffusive fluctuation}
\label{sec:teo}

The average shape of a fluctuation can be calculated through the
expression \cite{BaldassarriPRL03,ColaioriPRE}
\begin{equation}\label{xtT2NoMarkov} \langle
x(t)\rangle_T=\lim_{x_0\to 0^+} \overline{\frac{\int_0^\infty dx
\, x\,\Omega(x,t|\{x(0)\};x_0,0,x_0,T) } {\int_0^\infty dx \,
\Omega(x,t|\{x(0)\};x_0,0,x_0,T) }}
\end{equation}
where $\Omega(x,t|\{x(0)\};x_0,0,x_0,T)$ is the probability that
the walker with trajectory $\{x(0)\}$ for $t<0$ is at $x$ at time
$t$ provided that he was at $x_0>0$ at time $0$ and at $x_0$ at
time $T$ without ever touching the axis at $x=0$ in the time
interval $(0,T)$. The upper  line in Eq.\ \eqref{xtT2NoMarkov}
means average over all the trajectories $\{x(0)\}$ that reach
$x_0$ at time $t=0$. Let $F(x,t|\{x(t')\};x',t')$ be the
probability that the walker with trajectory $\{x(t')\}$ for $t<t'$
and who was at $x'>0$ at time $t'$ reaches $x>0$ at time $t$
without ever touching the axis at $x=0$; let
$\Omega(x,t|x_0,0,x_0,T)$ be the probability that the walker is at
$x$ at time $t$ provided that he was at $x_0>0$ at time $0$ and at
$x_0$ at time $T$ without ever touching the axis at $x=0$ in the
time interval $(0,T)$; and let $F(x,t|x',t')$ be the probability
that the walker who was at $x'$ at time $t'$ reaches $x>0$ at time
$t$ without ever touching the axis at $x=0$. For walks without
memory (Markovian walks)
$\Omega(x,t|\{x(0)\};x_0,0,x_0,T)=\Omega(x,t|x_0,0,x_0,T)$,
$F(x,t|\{x(t')\};x',t')=F(x,t|x',t')$, and one can write
$\Omega(x,t|x_0,0,x_0,T)$ as $F(x,t|x_0,0) F(x,T-t|x_0,0)$, so
that Eq.~\eqref{xtT2NoMarkov} becomes
\cite{BaldassarriPRL03,ColaioriPRE}
\begin{equation}\label{xtT2} \langle
x(t)\rangle_T=\lim_{x_0\to 0^+}\frac{\int_0^\infty dx \,
x\,F(x,t|x_0,0) F(x,T-t|x_0,0)} {\int_0^\infty dx \, F(x,t|x_0,0)
F(x,T-t|x_0,0) } \;.
\end{equation}
This equation \emph{is not exact} for subdiffusve walks because
they are not Markovian. However, for subdiffusive walks, the
memory, i.e., the effect of the fact that at $t'$ the particle was
at $x'$ on the probability that the particle  at time $t>t'$ is at
$x$, decays as $(t-t')^{-\gamma}/\Gamma(1-\gamma)$
\cite{MetKlaPhysRep}. This implies that the approximation of
$F(x,t|\{x(t')\};x',t')$ by $F(x,t|x',t')$, and, consequently, the
accuracy of Eq.~\eqref{xtT2} for subdiffusive walks, improves when
$t-t'$ increases and $\gamma$ is close to 1.

\begin{figure}
\includegraphics[width=0.77 \columnwidth]{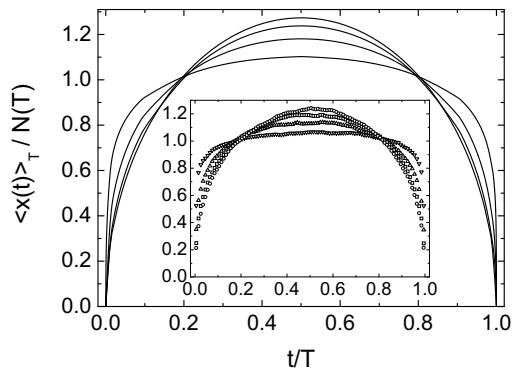}
\caption{Normalized average fluctuation for subdiffusive processes
for several values of $\gamma$. The shape is normalized so that
its area is 1, i.e., we plot $\langle x(t)\rangle_T/N(T)$. The
lines are the theoretical result for (at $t/T=1/2$ and from top to
bottom) $\gamma=1, 0.9,0.75,0.5$. Inset: simulation results for
these same values of $\gamma$. \label{xnteosimu.eps}}
\end{figure}

The probability $F(x,t|x_0,0)$ can be calculated by means of the
method of images $ F(x,t|x_0,0)=P(x-x_0,t)-P(-x-x_0,t)$
\cite{Rednerbook,MetKlaBoundary},  $P(x-x_0,t)$  being the
probability density that the free process (without boundary
conditions) that  at time $t\leq 0$ was at $x_0$ is  at $x$ at
time $t$. For subdiffusive processes, and for
$t^{-\gamma}/\Gamma(1-\gamma)\ll 1$ \cite{BarMetKlaPRE}, $P(x,t)$
can be written in terms of Fox's $H$ function as
\cite{MetKlaPhysRep}:
\begin{equation}\label{}
P(x,t)=\frac{1}{\sqrt{4 K_\gamma t^\gamma}}
H^{10}_{11}\left[\frac{|x|}{\sqrt{ K_\gamma t^\gamma}}
    \LM{(1-\gamma/2,\gamma/2)}{(0,1)}   \right].
\end{equation}
Taking into account that the Laplace transform of $P(x,t)$ is
\begin{equation}\label{}
P(x,u)=\frac{u^{\gamma/2-1}}{\sqrt{4K_\gamma}}
\exp(-\sqrt{u^\gamma/K_\gamma} |x|)
\end{equation}
one finds  for $x\ge 0$
\begin{align}\label{}
F(x,u|x_0,0)&=\frac{u^{\gamma/2-1}}{\sqrt{4K_\gamma}}  \left\{
e^{-a(x_0-x)} [\Theta(x)-\Theta(x-x_0)] \right. \nonumber \\
& \left. +e^{-a(x-x_0)}\Theta(x-x_0) - e^{-a(x+x_0)} \right\},
\end{align}
where $a\equiv \sqrt{u^\gamma/K_\gamma}$ and $\Theta(x)$ is the
Heaviside step function. As we are interested in the limit $x_0\to
0$ with $x
> x_0$,  we expand the term inside the bracket in powers of  $x_0$ and get
\begin{equation}\label{cxu} F(x,u|x_0\to 0,0)=x_0
\frac{u^{\gamma-1}}{K_\gamma} e^{-\sqrt{u^\gamma/K_\gamma}x }
\end{equation}
which implies
\begin{equation}\label{cxt}
F(x,t|x_0\to 0,0)=\frac{x_0 }{K_\gamma t^\gamma}
H^{10}_{11}\left[\frac{x}{\sqrt{ K_\gamma t^\gamma}}
    \LM{(1-\gamma,\gamma/2)}{(0,1)}   \right].
\end{equation}
Inserting this expression into Eq.\ \eqref{xtT2} and carrying out
the integrations \cite{MathaiSaxena} one finds that the average
shape of a subdiffusive stochastic process (subdiffusive random
walk) is given by
\begin{equation}
\label{xtT6} \langle x(t)\rangle_T=T^{\gamma/2}
f_\gamma(t/T)=\sqrt{K_\gamma T^\gamma} g_\gamma(t/T)
\end{equation}
 where
\begin{equation}
\label{fesca} g_\gamma(t/T)= \frac{
\left(\frac{t}{T}\right)^{\gamma/2}
H^{11}_{22}\left[\left(\frac{t/T}{1-t/T}\right)^{\gamma/2}
\LM{(-1,1),(1-\gamma,\frac{\gamma}{2})}{(0,1),(0,\frac{\gamma}{2})}
\right]}
{H^{11}_{22}\left[\left(\frac{t/T}{1-t/T}\right)^{\gamma/2}
\LM{(0,1),(1-\gamma,\frac{\gamma}{2})}{(0,1),(\frac{\gamma}{2},\frac{\gamma}{2})}
\right] }.
\end{equation}
In Fig.\ \ref{xnteosimu.eps} we  plot the (normalized) average
shape of fluctuations for several classes of subdiffusive
processes.  We see that the shape tends to a table-like form as
$\gamma$ decreases, i.e., as the subdiffusive character of the
process increases.

Of course, for $\gamma \to 1$ one recovers the Gaussian result $
f_1(t/T)=\sqrt{16 D/\pi} \sqrt{t/T (1-t/T)} $
\cite{BaldassarriPRL03,FisherJSP84} because the upper and lower
Fox's $H$ functions in Eq.\eqref{fesca} become $2\,z/\pi \,( 1 +
z^2) ^2$ and $z/2\,\sqrt{\pi }\,
    ( 1 + z^2)^{3/2}$, respectively.

The area $N(T)$ of the average fluctuation of duration $T$ is
given by
\begin{equation}\label{NgT0} N(T)=\int_0^1 ds \langle x(s
T)\rangle_T
\end{equation}
where $s=t/T$. From  Eq.~\eqref{xtT6} one finds
\begin{equation}\label{NgT}
N(T)=n_\gamma \sqrt{K_\gamma T^\gamma}
\end{equation}
where $n_\gamma=\int_0^1 ds g_\gamma(s)$.
Then, the normalized
average fluctuation is given by
\begin{equation}\label{}
\frac{\langle x(t)\rangle_T}{N(T)}=\frac{g_\gamma(t/T)}{n_\gamma}.
\end{equation}
We have not been able to calculate $n_\gamma$ analytically. In
Table \ref{tab:1} we give some values evaluated numerically.

\begin{table}
\caption{The coefficient  $n_\gamma=\int_0^1 ds g_\gamma(s)$
calculated  by numerical integration.}
\begin{ruledtabular}
\begin{tabular}{ccccccccc}
$\gamma$ &1/4&1/3&1/2&2/3& 3/4&4/5&9/10 &1 \\
$n_\gamma$ &0.493&0.501&0.537&0.612&0.668&0.708& 0.798& $\sqrt{\pi}/2$ \\
\end{tabular}
\end{ruledtabular}
 \label{tab:1}
\end{table}

\begin{figure}
\includegraphics[width=0.77 \columnwidth]{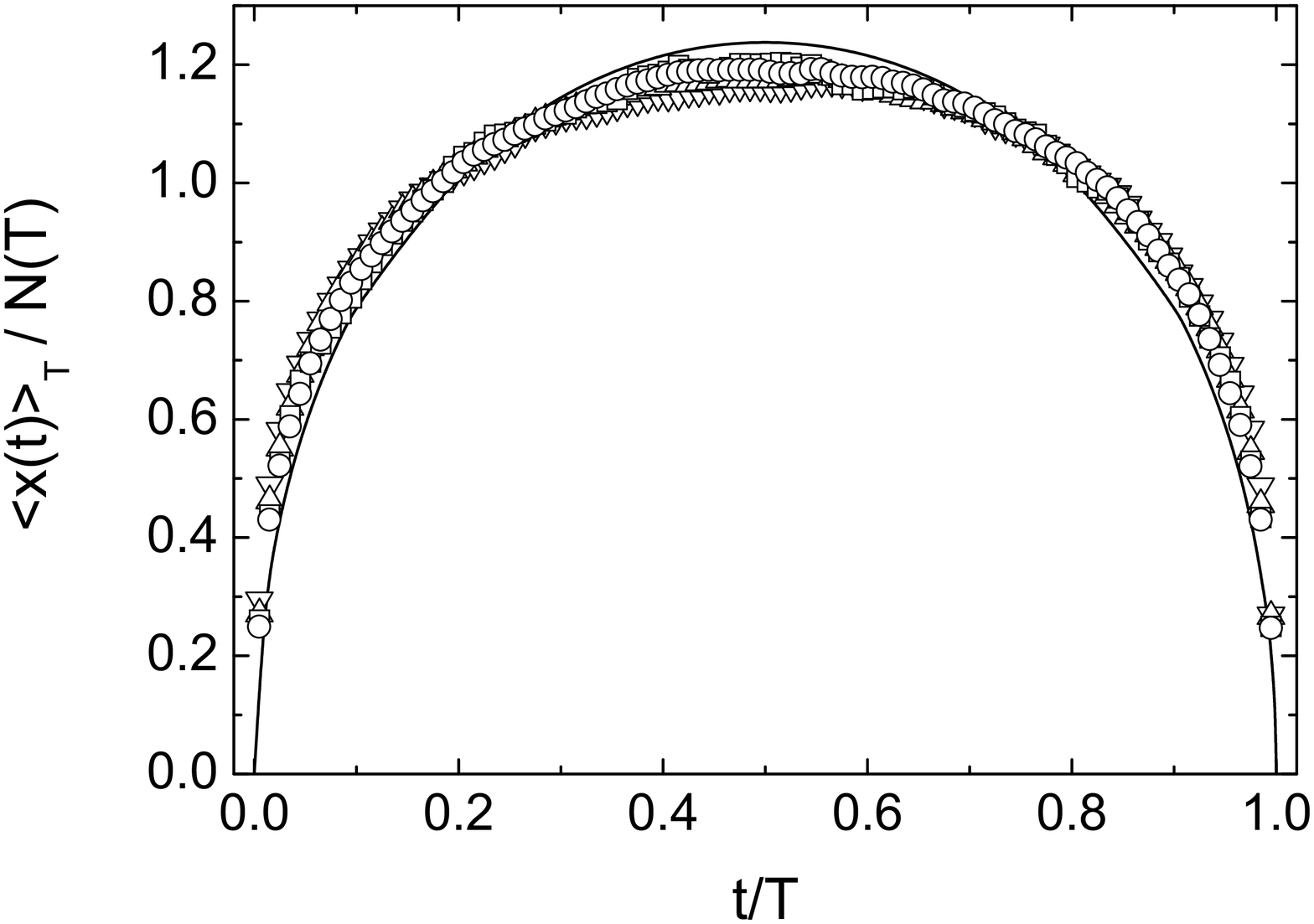}
\caption{Normalized average fluctuation for the subdiffusive process with
$\gamma=0.9$. The line is the theoretical result and the symbols are simulation
results for $T=10^4, 10^5, 10^6, 10^7, 10^8$. \label{shapega09.eps}}
\end{figure}

\begin{figure}
\includegraphics[width=0.77 \columnwidth]{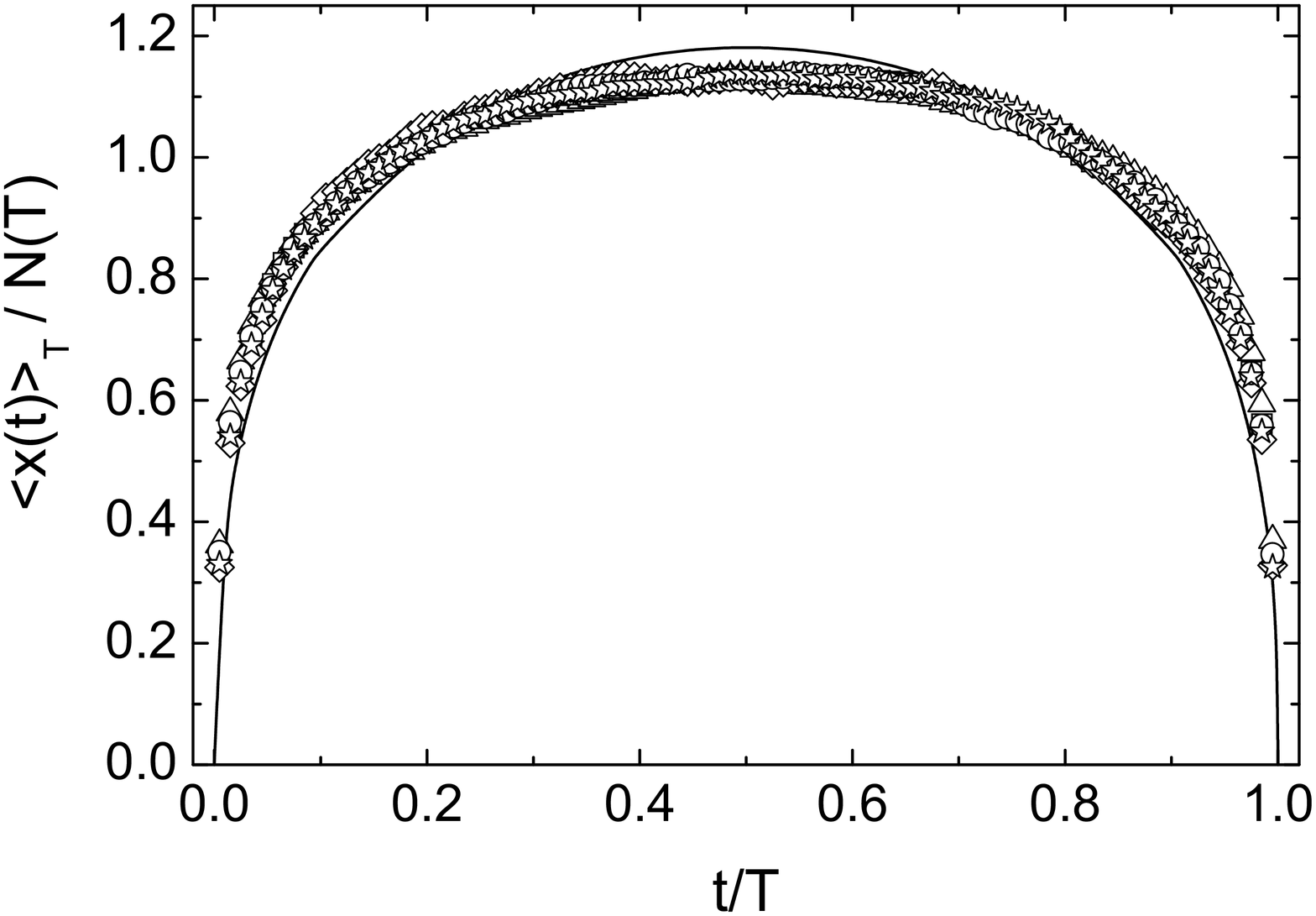}
\caption{Normalized average fluctuation for the subdiffusive process with
$\gamma=0.75$. The line is the theoretical result and the symbols are
simulation results for $T=10^4, 10^5, 10^6, 10^7, 10^8$.
\label{shapega075.eps}}
\end{figure}

\begin{figure}
\includegraphics[width=0.77 \columnwidth]{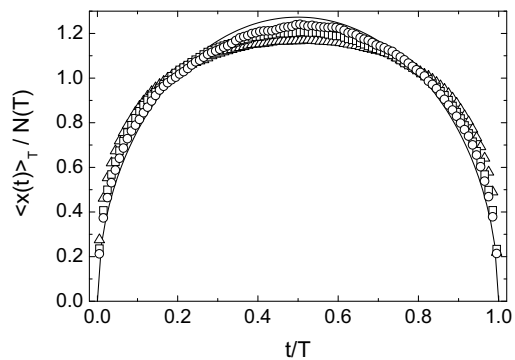}
\caption{Normalized average fluctuation for a process with $\gamma=1$. The line
is the semicircular form theoretical result and the symbols are simulation
results for $T=10^3$ (triangles), $10^5$ (squares), and $10^7$ (circles).
\label{xnga1.eps}}
\end{figure}

\section{Simulation}
\label{sec:simu}

We carried out simulations of the fluctuations of the stochastic
process \eqref{xtstosub} where $\delta x$ takes the values $+1$ or
$-1$ with equal probabilities  and where the waiting time  $\delta
t$ between jumps follows the Pareto distribution
$\psi(t)=\gamma/(1+t)^{1+\gamma}$. In this case, the diffusion
constant $K_\gamma$ is given by \cite{MetKlaPhysRep}
$K_\gamma=1/[2\Gamma(1-\gamma)]$. The simulation results follow
the pattern found in the precedent section: the fluctuation shape
tends toward a table-like form as $\gamma$ decreases (see inset in
Fig. \ref{xnteosimu.eps}).

In Figs.\  \ref{shapega09.eps} and \ref{shapega075.eps} we compare the
theoretical predictions and the simulation results for $\gamma=0.9$ and
$\gamma=0.75$, respectively. The agreement is reasonable. We attribute the
differences to, first, the approximate nature of Eq.~\eqref{xtT2} for
non-Markovian processes and, second, to a slow convergence that requires longer
$T$ to set in. This can be clearly appreciated in Fig.\ \ref{xnga1.eps} where
one sees that the well-established theoretical semicircular shape is
approached, although very gradually, as $T$ increases . It is even more gradual
for smaller $\gamma$. We have not explored larger values for $T$ because of the
excessive computer time required.

\begin{figure}
\includegraphics[width=0.77 \columnwidth]{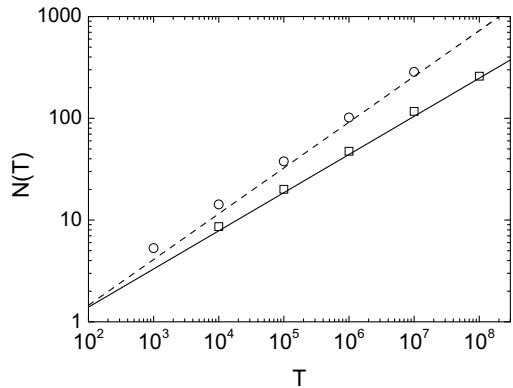}
\caption{Area $N(T)$ of the fluctuation  for $\gamma=0.75$
(circles) and $\gamma=0.9$ (squares). The lines are the
corresponding theoretical prediction; see
Eq.~\protect{\eqref{NgT}.} \label{NT.eps}}
\end{figure}

In the simulations we also calculated the area $N(T)$ of the
fluctuation. In Fig.~\ref{NT.eps} we plot $N(T)$ for several
values of $T$ and for two values of $\gamma$. The agreement
between simulation and theory is reasonable again.

\section{Conclusions}
We have analyzed the average shape of the fluctuations of time
series generated by the stochastic process \eqref{xtstosub} with a
power-law waiting time distribution $\psi(t)\sim t^{-1-\gamma}$
where $0<\gamma<1$ (subdiffusive walks). Although the spatial
increments $\delta x$ in this stochastic process are uncorrelated,
we find that, in contrast with Gaussian walks and L\'evy flights,
the average shape of a fluctuation is no longer a semicircle: its
form becomes flatter at the top and  steeper at the extremes as
the subdiffusive character of the process increases (i.e., as
$\gamma$ decreases).

Some  possible sequels of the present work are obvious: one could consider the
effect of correlated increments $\delta x$ (as done by Baldassarri \textit{et
al.} \cite{BaldassarriPRL03}) or investigate the average shape of a fluctuation
for L\'evy walks (in which $|\delta x|$ and $\delta t$ are proportional and
follow a broad power-law tail). However, for these cases,  the task of getting
analytical results, even approximate, such as those reported in the present
paper certainly looks formidable.

\acknowledgments We are grateful to Francesca Colaiori, Andrea Baldassarri and
Claudio Castellano for their useful remarks and for sharing with us Ref.\
\cite{ColaioriPRE} prior to publication. This work was supported by the
Ministerio de Ciencia y Tecnolog\'{\i}a (Spain) through Grant No. FIS2004-01399
and by the European Community's Human Potential Programme under contract
HPRN-CT-2002-00307, DYGLAGEMEM.



\begin{thebibliography}{99}

\bibitem{FractalBooks}
H. Takayasu, {\em Fractals in the Physical Sciences} (Wiley,
Chichester, 1992); {\em Fractals in Science}, edited by A. Bunde
and S. Havlin (Springer-Verlag, Berlin, 1994); M. Schroeder, {\em
Fractals, Chaos, Power Laws: Minutes from an Infinity Paradise}
(Freeman, New York, 1991).


\bibitem{BouchaudCM00} J.-P. Bouchaud, e-print cond-mat/0008103, http://xxx.lanl.gov


\bibitem{BakPhysicaA97} P. Bak, M. Paczuski and M. Shubik, Physica
A \textbf{246}, 430 (1997).

\bibitem{FederBookFractals} J. Feder, {\em Fractals} (Plenum, New York, 1988).

\bibitem{KuntzSethnaPRB00} M. C. Kuntz and J. P. Sethna, Phys.
Rev. B \textbf{62}, 11699 (2000).


\bibitem{SpasojevicPRE96} D. Spasojevi\'c, S. Bukvi\'c, S. Milo\v{s}evi\'c
and H. E. Stanley, Phys. Rev. E \textbf{54}, 2531 (1996).

\bibitem{BouchaudJPI} J.-P. Bouchaud, J. Phys. I \textbf{2},  1705 (1992).

\bibitem{ArousPRL02} G. Ben Arous, A. Bovier and V. Gayrard, Phys. Rev. Lett. \textbf{88}, 087201
(2002).
\bibitem{BertinPRE03} E. M. Bertin and J.-P. Bouchaud, Phys. Rev.
E \textbf{67}, 026128 (2003).


\bibitem{SaubameaPRL99} B. Saubam\'ea, M. Leduc and C.
Cohen-Tannoudji, Phys. Rev. Lett. \textbf{83}, 3796 (1999).

\bibitem{ShimizuPRB01} K. T. Shimizu, R. G. Neuhauser, C. A.
Leatherdale, S. A. Empedocles, W. K. Woo and M. G. Bawendi, Phys.
Rev. B \textbf{63}, 205316 (2001).


\bibitem{BrokmannConMat02} X. Brokmann, J.-P. Hermier, G. Messin,
P. Desbiolles, J.-P. Bouchaud and M. Dahan, e-print cond-mat/0211171,
http://xxx.lanl.gov

\bibitem{MetKlaPhysRep}
R. Metzler and J. Klafter, Phys. Rep. {\bf 339}, 1 (2000).

\bibitem{BouchaudPhysRep90} J.-P. Bouchaud and A. Georges, Phys.
Rep. \textbf{195}, 127 (1990).


\bibitem{HavlinAvrahamAdvPhys87} S. Havlin and D. Ben-Avraham,  Adv. Phys.  {\bf 36},  695 (1987).

\bibitem{BaldassarriPRL03} A. Baldassarri, F. Colaiori and C.
Castellano, Phys. Rev. Lett. {\bf 90},  060601 (2003).

\bibitem{ColaioriPRE}  F. Colaiori , A. Baldassarri and C.
Castellano, e-print cond-mat/0402285, http://xxx.lanl.gov


\bibitem{SethnaNature01} J. P. Sethna, K. A. Dahmen and C. R. Myers, Nature \textbf{410}, 242 (2001).

\bibitem{FisherJSP84} M. Fisher, J. Stat. Phys. \textbf{34}, 667
(1984).

\bibitem{Rednerbook}
S. Redner, {\em  A Guide to First-passage Processes} (Cambridge University
Press, Cambridge, 2001).

\bibitem{MetKlaBoundary} R. Metzler and J. Klafter, Physica A \textbf{278}, 107 (2000).

\bibitem{BarMetKlaPRE} E. Barkai, R. Metzler and J. Klafter,
Phys. Rev. E {\bf 61}, 132 (2000).

\bibitem{MathaiSaxena} A. M. Mathai and R. K. Saxena,
{\em The H-function with Applications in Statistics and Other
Disciplines} (Wiley, New York, 1978).


\end{thebibliography}
\end{document}